\documentclass[preprint,prd,nofootinbib,onecolumn,showpacs]{revtex4}
\usepackage{dcolumn}
\usepackage{lipsum}
\usepackage{graphicx,epsfig}
\usepackage{psfrag}
\usepackage{bm}
\usepackage{epstopdf}
\usepackage{latexsym}
\usepackage{multirow}
\usepackage{amsmath,amssymb}
\usepackage{enumerate}

\def\be {\begin{equation}}
\def\ee {\end{equation}}
\def\ba {\begin{eqnarray}}
\def\ea {\end{eqnarray}}
\def\nn {\nonumber}
\def\bc {\begin{center}}
\def\ec {\end{center}}
\def\b  {\beta}
\def\c  {\gamma}

\def\e  {\epsilon}

\newcommand{\bdm}{\begin{displaymath}}
\newcommand{\edm}{\end{displaymath}}

\def\k  {\kappa}

\def\O  {\Omega}
\def\p  {\pi}

\def\r  {\rho}

\def\t  {\tau}
\def\la {\label}
\def\le {\left}
\def\ri {\right}
\def\pa {\partial}
\def\f {\frac}

\def\bi {\begin{itemize}}
\def\ei {\end{itemize}}

\def\bc {\begin{center}}
\def\ec {\end{center}}
\def\vph {\varphi}

\begin{document}

\title{Entanglement signatures of phase transition in higher-derivative quantum field theories} 

\author{Suman Ghosh} \email[email: ]{suman.ghosh@iisertvm.ac.in}
\author{S. Shankaranarayanan} \email[email: ]{shanki@iisertvm.ac.in}
\affiliation{School of Physics, Indian Institute of Science Education and 
Research, Thiruvananthapuram 695016, India}

\begin{abstract}
We show that the variation of the ground state entanglement in 
linear, higher spatial derivatives field theories at zero-temperature 
have signatures of phase transition. 
Around the critical point, when the dispersion relation changes from linear to non-linear, there 
  is a fundamental change in the reduced density matrix leading to a change
  in the scaling of entanglement entropy. We suggest possible
explanations involving both kinematical and dynamical effects. 
  We discuss the implication of our work for 2-D condensed matter 
  systems, black-hole entropy and models of quantum gravity. 
\end{abstract}

\maketitle

\begin{center}
\section{Introduction}
\end{center}

A composite quantum system can be prepared in the so-called {\it
  Entangled} (inseparable) states. These states precisely measure the
non-local quantum correlations between the subsystems which can not be
explained by classical physics. Entanglement is at the heart of some
of the deepest puzzles of quantum mechanics and its most promising
applications \cite{2009-Horodecki.etal-RMP,2010-Eisert.etal-RMP}.

In recent years, entanglement entropy is found to be playing crucial
roles in understanding quantum behavior of macroscopic and microscopic
systems.  For the macroscopic systems like black-holes, entanglement
entropy refers to the measure of the information loss (for an outside
observer) due to the spatial separation between the degrees of freedom
inside and outside the horizon. Interestingly, the entanglement
entropy turns out to be proportional to the area of the horizon
\cite{1986-Bombelli.etal-PRD,1993-Srednicki-PRL,2008-Das.etal-PRD,2010-Eisert.etal-RMP}
and raises the possibility to interpret the Bekenstein-Hawking entropy
as the entanglement entropy.

At the microscopic level, it has been observed that the entanglement
plays an important role in quantum phase transitions
~\cite{1997-Sondhi.etal-RMP,2001-05-30-SachdevSubir-Quantumphasetransitions,2011-Carr-QPT}.
Typically at the critical point, where the transition takes place,
long-range correlations in the ground state develop. Like the free
energy for the ordinary phase transitions, entanglement is an useful
quantifying measure to understand the long-range correlations in
quantum phase
transitions~\cite{2002-Osterloh.etal-Nature,2002-Osborne.Nielsen-PRA,2004-Wu.etal-PRL,2010-Rieper.etal-NJP,1972-Anderson-Science}.
The simplest versions of phase-transitions assume the (scalar) order
parameter to have non-linear self-interactions $ F(\phi) = a_2 \phi^2
+ a_4 \phi^4 + \cdots$ \cite{1987-Toledano-Landau}. One way to
generalize the above model is to include inhomogeneities in the
order-parameter of the form \cite{1975-Honreich.etal-PRL}: 
{\small \be
  F(\phi) = \sum_{n = 1}^{\infty} a_{2 n} \phi^{2 n} + \sum_{m =
    1}^{\infty} b_m \left( \nabla^m \phi \right)^2 \, .  \ee }
We show that a fundamental change in the ground state of a quantum
system arises due to higher spatial derivative terms (especially, $b_3
\neq 0$) even if the self-interactions are set to zero ($a_{2n} =
0$). In particular, we show that for a real scalar field $(\phi)$
whose Hamiltonian
is~\cite{1995-Unruh-PRD,1996-Corley.Jacobson-PRD,1999-Padmanabhan-PRD,2009-Visser-PRD}:\\
{\small
\be
H = \f{1}{2} \int d^3x \le[ \p^2(x) + \sum_{m = 1}^{\infty} b_{m}
\left({\nabla}^{m}\phi\right)^2\ri] \, ,
\label{eq:Ham}
\ee
}
the entanglement entropy, computed for the discretized Hamiltonian, shows a sudden discontinuity 
due to odd coupling coefficients $(b_m)$\footnote{Such Hamiltonians are commonly
  encountered in liquid crystals and strain
  dynamics~\cite{2003-Lookman.etal-PRB} by rewriting the above
  Hamiltonian in terms of the strain vector $e_{i} = \nabla_i \phi$.}.
We explicitly show this up to $m = 3$. 
In the above Hamiltonian, we let $b_m = \pm \k^{-2(m-1)}$ where $\k$ is the 
spatial frequency at which the dispersion relation changes from linear to non-linear.  
In the next section, we briefly discuss the field theory model, we consider, and the computational method to determine the entanglement entropy.

\section{The field theory model}

Our model is the Hamiltonian (\ref{eq:Ham}) which corresponds to a
real scalar field propagating in $(3+1)$-dimensional flat space-time
with higher spatial derivative terms at zero temperature. To extract
the physics, we will consider the following dispersion relation:
{\small
\be 
\omega^2 = k^2 + \f{\epsilon}{\kappa^2} k^4 + \f{\tau}{\kappa^4} k^6 \, ,
\label{eq:dispersion2}
\ee}
keeping the terms up to $m = 3$ in Hamiltonian (\ref{eq:Ham}), where
$\epsilon, \tau$ are (dimensionless) constants.

It is harder to obtain an analytical expression for entanglement for
free fields \cite{1986-Bombelli.etal-PRD,1993-Srednicki-PRL}.  In this
work, we use the following semi-analytic approach: \\
(a) Discretize the Hamiltonian (\ref{eq:Ham}) along the radial
direction of a spherical lattice with lattice spacing $a$, such that
$r \rightarrow r_i = ia;~r_{i+1}-r_i=a$:
{\small
\be
H = \sum_{j} H_{j} = \frac{1}{2a} \sum_{i,j}^{N} \delta_{ij} \pi_{j}^2 + \vph_{j} \, K_{ij} \, \vph_i \, .
\label{eq:discretizeHam}
\ee } 
Note that $\vph_i$'s, $\pi_i$'s and $K_{ij}$'s are dimensionless \cite{1993-Srednicki-PRL} and the 
interactions are contained in the off-diagonal elements of the matrix $K_{ij}$ (see Appendix A) that are functions of 
$P,\ell,N,\e$ and $\t$\footnote{Note that $K_{ij}$ contains
nearest-neighbor (nn), next to nn (nnn), next to nnn (nnnn) and next
to nnnn (nnnnn) interaction terms which originate in the derivative
terms in Eq.(\ref{eq:Ham}). The central difference scheme we
  use is superior compared to the mid-point discretization scheme in
  the literature \cite{1993-Srednicki-PRL}.}. Here $P= \f{1}{a^2\k^2}$ and $P^{m-1}$ is the (dimensionless) effective 
coupling coefficient corresponding to the $m$-th term in Eq. (\ref{eq:discretizeHam}).
Using the fact that the discretization scale $a$ should be greater
than the cutoff scale ($1/\kappa$), we have $0 \leq P \leq 1$.\\
(b) The lattice is terminated at a large but finite $N$. An
intermediate point $n$ is chosen, which we call the {\it horizon}
${\cal R}$ ($= an$), that separates the lattice points between the
{\it inside} and {\it outside}.  The horizon acts as a boundary of the
bipartite system.
(c) We confine our interest to the entanglement between the degrees of
freedom inside and outside the horizon. The reduced density matrix
$\rho_{_{\rm reduced}}$ (see Appendix A) is obtained by tracing over all the lattice points
inside the horizon for the ground state wave-function of the
discretized Hamiltonian (\ref{eq:discretizeHam}). The resulting
$\rho_{_{\rm reduced}}$ is a mixed state of a bipartite system. Entanglement is
computed as the von Neumann entropy associated with the reduced
density matrix $\rho_{_{\rm reduced}}$ \cite{2009-Horodecki.etal-RMP,2010-Eisert.etal-RMP}:
\be
S^{(P)} = \mbox{Tr} \, \left( \rho_{_{\rm reduced}} \ln \rho_{_{\rm reduced}} \right)
\ee
In the next section, we show the results of numerical computations of entanglement entropy (following the method described in Appendix A) for our model.

\section{Numerical results}

\begin{figure}[!htb]
\begin{center}\hspace*{-24pt} 
\includegraphics[scale=0.4]{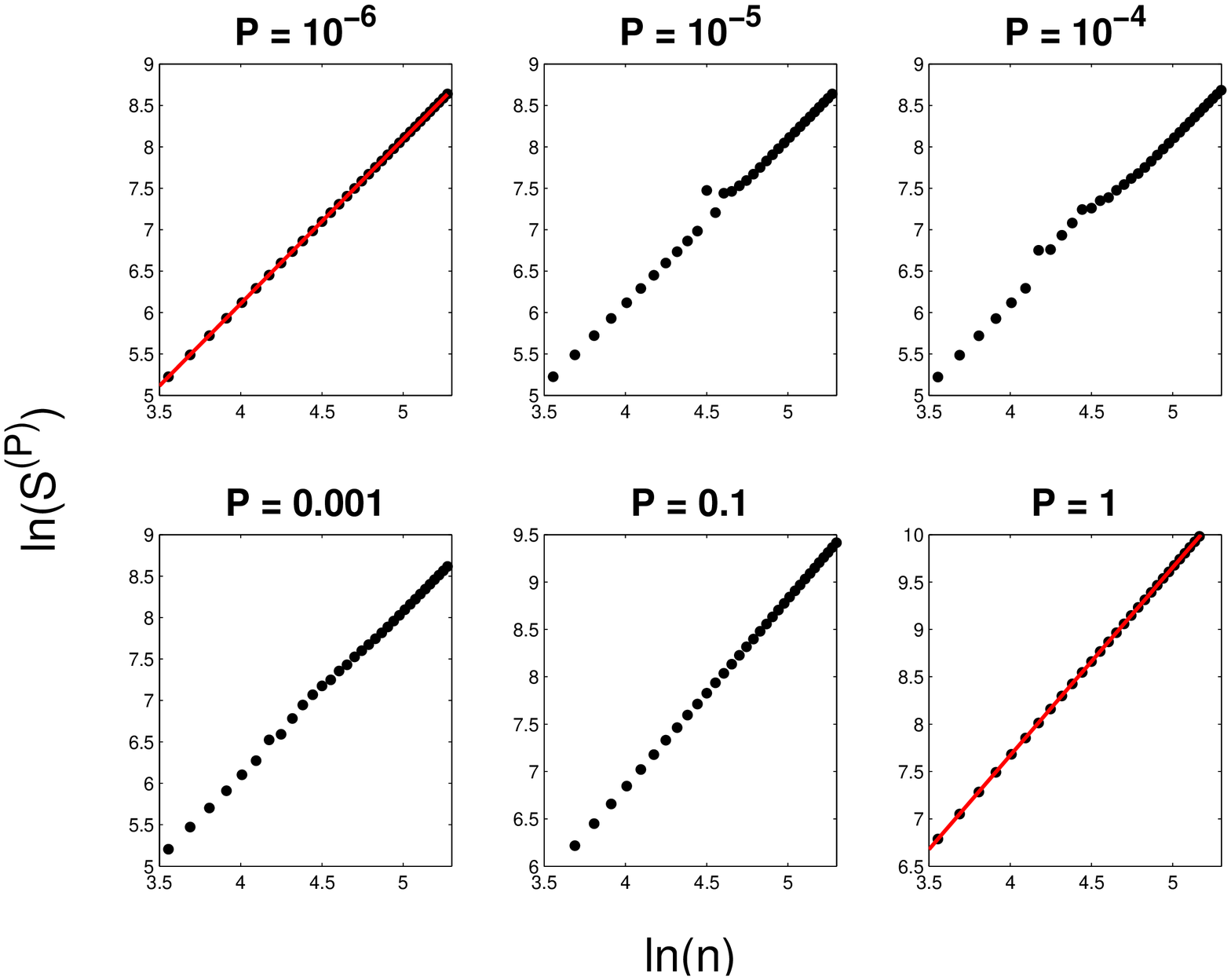}
\caption{Log-log plot of von Neumann entropy, S, versus the scaled
  radius of the sphere ${\cal R}/a = n$, for $\tau =0$, $\epsilon = 1$
  and different values of coupling parameter $P$. The lattice size is
  $N = 300$ and $50 \leq n \leq 290$. The dots represent the numerical
  output and solid lines denote lines of best
  fit.}\label{fig:2ndorder}
\end{center}
\end{figure}

We compute the entanglement entropy for the discretized Hamiltonian
(\ref{eq:discretizeHam}) for different coupling strength $P$ (Appendix B proves convergence of total entropy as $\ell\rightarrow\infty$).  The
computations are done using Matlab for the lattice size $N = 300, 50
\leq n \leq 295$ and the relative error in the computation of the
entropy is $10^{-5}$.  The computations are done for the following
three different scenarios:

{\large{\tt (1)}} $\tau =0$, $\epsilon = 1$: In
Fig.~(\ref{fig:2ndorder}), we have plotted $\ln S^{(P)}$ versus
$\ln({\cal R}/a)$ for different value of the coupling parameter $P$.
The best fit (solid) curves in the two asymptotic regimes,
$P\rightarrow 0$ and $P\rightarrow 1$, show that the entropy scales
approximately as area $S \sim ({\cal R}/a)^2$.  As we increase $P$,
however, the prefactor increases.  For $P = 10^{-5}$ and $10^{-4}$,
around the transition region between linear to non-linear, entropy
increases by an order.

\begin{figure}[!htb]
\begin{center}
\hspace*{-22pt} \includegraphics[scale=0.4]{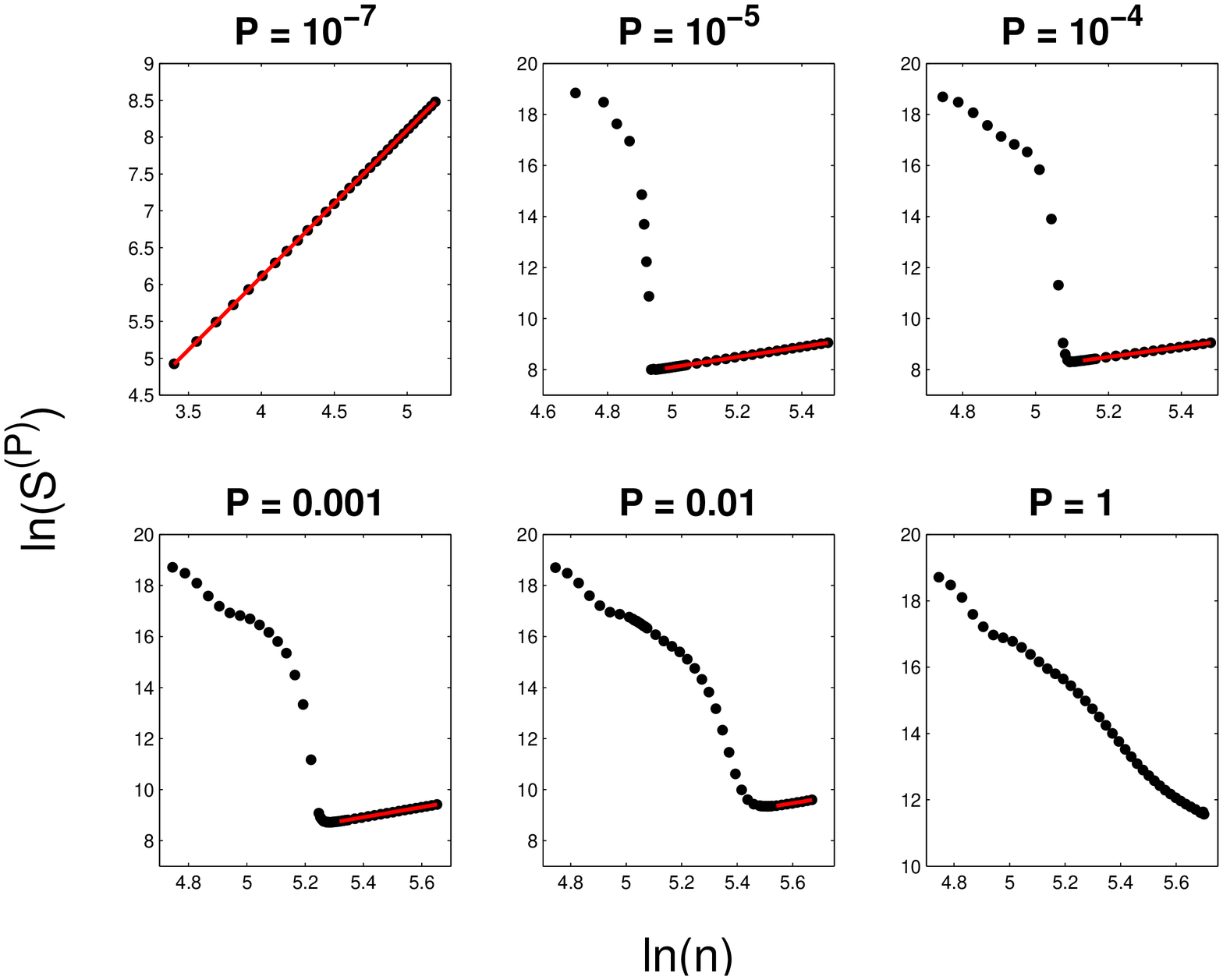} 
\caption{Log-log plot of von Neumann entropy, S, versus the scaled
  radius of the sphere ${\cal R}/a = n $, for $\tau =1$, $\epsilon =
  0$ and different values of coupling parameter $P$. The lattice size
  is $N = 300$ and $100 \leq n \leq 290$. $n_c$ for $P = 10^{-5},
  10^{-4}, 10^{-3}$, respectively, are 139, 167 and 197.}
\label{fig:3rdorder}
\end{center}
\end{figure}

{\large{\tt (2)}} $\tau =1$, $\epsilon = 0$: In
Fig.~(\ref{fig:3rdorder}), $\ln S^{(P)}$ versus $\ln({\cal R}/a)$ are
plotted for different coupling parameter $P$, from which we infer the
following: (i) For small $P$, as in the earlier case, entropy scales
as area.
(ii) As we increase $P$, above a \textit{critical} value of $P$, say
$P_c$ (here $P_c \sim 10^{-6}$) a new phase appears in the $\log-\log$
plots of entropy vs $n$.
(iii) For a given $P>P_c$, scaling of $S^{(P)}$ changes drastically
across some \textit{critical} value of $n$, say $n_c$ (e.g. $n_c \sim
139$ for $P=10^{-5}$).  For $n > n_c$, $S^{(P)}$ increases with $n$
and scales approximately close to area. However, for $n < n_c$,
$S^{(P)}$ increases with decreasing $n$ i.e.  $S^{(P)} \propto
n^{-\alpha}$ ($\alpha > 0$)\footnote{Note that as $n\rightarrow 0$,
  the entropy becomes zero. This is expected as the number of degrees
  of freedom gradually vanishes.}
(iv) With increasing $P$, $n_c$ increases.
(V) Near $P=P_c$ transition is sharp. For $P > P_c$, the transition
region is wider.

\begin{figure}[!htb]
\begin{center}
\hspace*{-18pt} \includegraphics[scale=0.4]{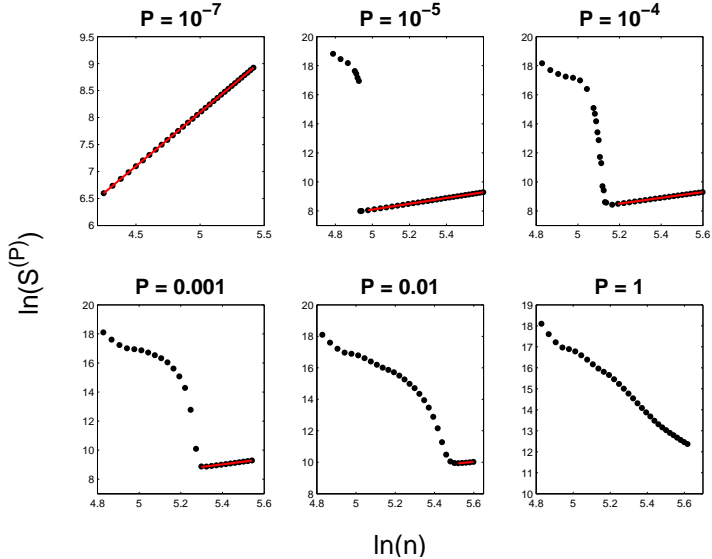} 
\caption{Log-log plot of von Neumann entropy, S, versus the scaled
  radius of the sphere ${\cal R}/a = n$, for $\tau =1$, $\epsilon =
  -1$ and different values of coupling parameter $P$. The lattice size
  is $N = 300$ and $100 \leq n \leq 290$. The dots represent the
  numerical output and solid lines denote lines of best fit.}
\label{fig:2nd&3rdorder}
\end{center}
\end{figure}

{\large{\tt (3)}} $\tau =1$, $\epsilon = -1$: Fig. (\ref{fig:2nd&3rdorder}) also shows similar
cross-over of entanglement entropy as in the earlier case.

\section{understanding the results}

The above numerical results clearly indicate that higher derivative
terms seem to drastically modify the vacuum structure, due to which
the entropy jumps by few orders of magnitude close to the critical
point, and throws up a large number of interesting questions. In the
rest of this article, we address three key questions related to this
new phenomena:
\begin{enumerate}[(I)]
\item What universal feature the higher derivative terms have on the
  scaling of the entropy?
\item What causes the entropy to increase by couple of orders of
  magnitude at the cross-over?
\item Does the sharp jump indicate phase transition?
\end{enumerate}

\subsection{Inverse scaling}

\begin{figure}[!htb]
\begin{center}
\includegraphics[scale=0.35]{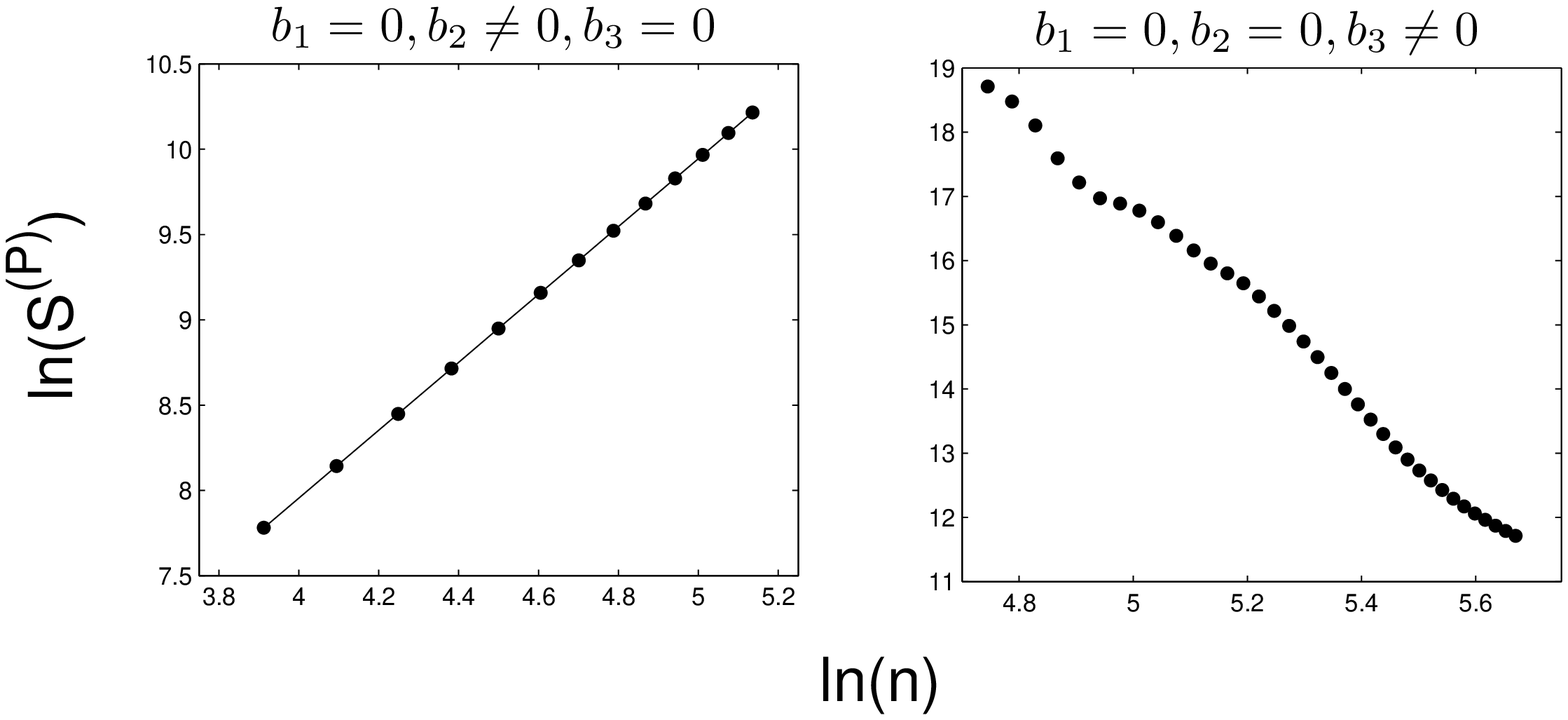} 
\caption{Log-Log plot of the entropy variation with only the 2nd order derivative term 
(left) and 3rd order derivative term (right) present in the Hamiltonian.}
\label{fig:only2ndand3rdorder}
\end{center}
\end{figure}
To go about answering question (I), let us set the dispersion relation
for all $k$ modes to be $\omega^2 = k^4/\kappa^2$ and $\omega^2 = k^6/\kappa^4$.  
Repeating the earlier analysis, $\ln S^{(P)}$ vs $\log n$ in
Fig. (\ref{fig:only2ndand3rdorder}) shows that the fourth order derivative
term leads to the usual area law while the sixth order derivative term 
shows inverse scaling i.e. entropy increases as we decrease $n$.  Comparing Fig. 
(\ref{fig:only2ndand3rdorder}) with $P = 1$ plots in Fig. (\ref{fig:3rdorder}) 
show that increasing $P$ leads to the dominance of the higher-order derivative
terms implying that the inverse scaling of entropy {\it uniquely}
corresponds to the presence of higher (sixth) order derivative terms.
The system has two distinct entropy profiles, the area law 
due to $(\nabla \phi)^2$ and inverse scaling due to $(\nabla^3 \phi)^2$. 
Hence, for a particular coupling strength $P^2$, the two distinct entropy
profiles appear at two regimes and a crossover happens at a $n_c$ that 
depends on $P$.

So, why the sixth-order derivative term leads to inverse scaling while the 
fourth-order derivative does not? One could plausibly get a better understanding 
of the phenomena if we look at the density of states and the two-point correlation
function  (Appendix C) for the three dispersion relations as given in Table (\ref{table1}). 
The density of states for the two dispersion relations ($k^2$ and $k^4$) increase
with energy while for $\omega^2 \propto k^6$, it is constant. The density of 
states is a kinematical quantity, the two point-function (See Appendix C)
contains information about the quantum fluctuations. The two point function is 
scale-invariant for the dispersion relation  $\omega^2 \propto k^6$. 
We discuss the implications for condensed matter systems in conclusions.

%

%
\begin{widetext}
\begin{table}[!h]
\begin{tabular}{|c|cc|cc|}
\hline 
Dispersion relation & & Density of states & & Two-point function  \\
\hline
$\omega^2 \propto k^2$ & &  $D(\omega) d\omega \propto  \omega^2 d\omega$ & & $G^+ \propto \f{1}{r^2} $  \\
\hline
$\omega^2 \propto k^4$ & & $D(\omega) d\omega \propto  \omega^{1/2} d\omega$  
&  & $   G^+ \propto \f{1}{r}$\\
\hline
$\omega^2 \propto k^6$ & & $D(\omega) d\omega \propto  {\rm constant}~d\omega$ 
& & Scale-invariant $ $ \\
\hline
\end{tabular}
\caption{Behavior of the density of states and two-point function for three dispersion
relations: (i) $b_1 \neq 0, b_2 = b_3 = 0$, (ii) $b_2 \neq 0, b_1 = b_3 = 0$, (iii) $b_3 \neq 0, b_1 = b_2 = 0$.}
\label{table1}
\end{table}
\end{widetext}
%

\subsection{Critical mode}
The plot of the distribution of entropy per partial wave $(2 \ell + 1)
S_{\ell}$ versus $\ell$ can be used to answer question (II).
Fig.~(\ref{fig:population}) shows the entropy distribution for three
values of $n$ for $\epsilon = 0, \tau = 1$ and $P = 10^{-5}$ --- 
$n = 200$ (which falls in the positive slope region in Fig. 
\ref{fig:3rdorder}), $n = 137$ (which falls on the
inverse scaling region in Fig. \ref{fig:3rdorder}) and $n=110$.  
It can be seen that, for $n = 200$, the contribution of large 
$\ell$ falls off rapidly implying that only lower partial waves 
$\ell$ contribute to the entropy. However, for $n = 137$, at a 
critical mode ($\ell_c$), higher partial waves contribute significantly 
to entropy compared to lower $\ell$. For $n=110$ this happens at a lower $\ell_c$.
This implies that, for a fixed $P$, as $n$ decreases, the turn around at 
which higher value of $\ell$ contribution to the entropy occurs earlier 
which in-turn leads to inverse scaling of entropy. This feature can be 
derived analytically by looking at $\ell \gg N$ limit of Hamiltonian 
(\ref{eq:discretizeHam}): 

\begin{figure}[!h]
\begin{center}
\includegraphics[scale=0.4]{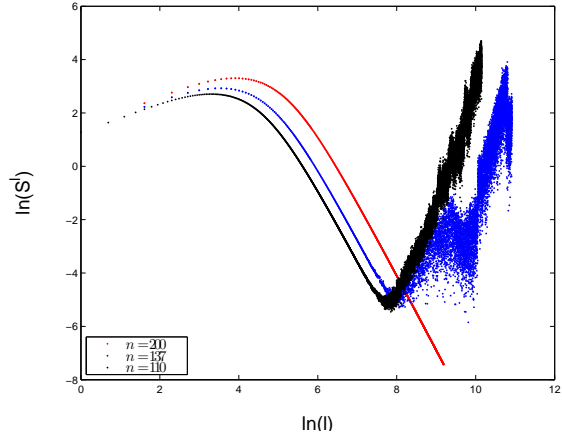}
\caption{Log-Log plot of the entropy distribution per partial wave $S^{\ell}=(2
  \ell + 1) S_{\ell}$ and $\ell$ for $n = 200$, $137$ and $110$ for $P =
  10^{-5}$. 
  }
\label{fig:population}
\end{center}
\end{figure}
\be
H \sim \f{1}{2a} \sum_{i} \pi_i^2 + \f{l^2}{i^2} \left(1 +  \tau P^2 ~\f{l^4}{i^4}\right)\phi_i^2 + 
\mbox{higher-order terms}
\ee
In the region when the higher-derivative term dominates, we have,
\be
\ell \ge P^{-1/2} {n}  \, .
\label{eq:p-n-l_c}
\ee
For a fixed $P < 1$, as seen in Fig. (\ref{fig:population}), as $n$ increases the 
cross over occurs at larger $\ell$. In the case of $P \sim 1$, this implies that 
the cross over occurs much earlier leading to inverse scaling of entropy\footnote{Eq. (\ref{eq:p-n-l_c}) apparently 
implies that inverse scaling phase in entropy profiles should appear for any arbitrary low value for 
$P>(n/\ell_c)^2$ . However, for very large $\ell$ the entropic contribution from each $\ell$ is 
already suppressed and for {\it low enough} $P$, $\ell_c$ will become so large that contributions 
from $\ell>\ell_c$-modes will be too small to perturb the area scaling.}. For a detailed comparison of the entropy per mode among systems with different dispersion relations, see Appendix D.
A qualitative understanding of excitation of higher modes (for $P>P_c$) can be understood 
using the quantum mechanical model of a particle in a box (see Appendix E). 

\subsection{The cross-over}

\subsubsection{Thermodynamic limit}
To answer question (III) that the change in the ground state signals
phase transition, we first calculate the entanglement entropy for different
values of $N$ with $P = 10^{-5}, \tau = 1, \epsilon = 0$ and accuracy
$10^{-5}$ (see Appendix F).
Fig. \ref{fig:n_c/N} clearly shows the ratio $n_c/N$ remains almost
constant with increasing $N$, implying that, the cross-over is indeed
a quantum phase transition.  It is important to note there is a no
phase transition for $N = 75$ indicating that {\it $P_c$ decreases
with increasing $N$}.
\begin{figure}[!h]
\begin{center}
\includegraphics[scale=0.45]{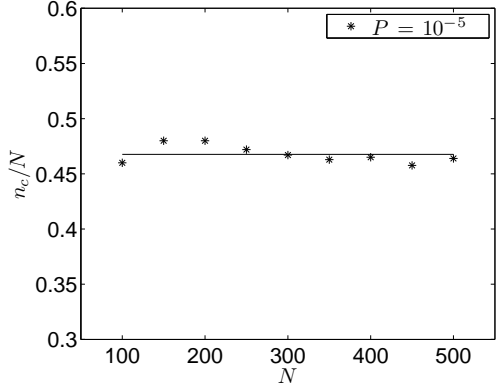}
\caption{Variation of $\f{n_c}{N}$ vs $N$ for $P=10^{-5}$.}
\label{fig:n_c/N}
\end{center}
\end{figure}

\subsubsection{Scaling transformations}
If $P_c$ is the quantum critical point, from the Wilsonian 
view point, at this point the (quantum) fluctuations occur at all length scales 
implying that action corresponding to Hamiltonian (\ref{eq:Ham}) or (\ref{eq:discretizeHam}) 
is scale invariant. Under the scaling transformations $(e^{-l})$,
\ba
x_i' &=& x ~ e^{-l} ;~~ i = 1,2,3\\
t' &=& t ~e^{-l},\\
\phi' &=& \phi ~e^l , \\
\k' &=& \k ~e^l.
\ea
the action corresponding to (\ref{eq:discretizeHam}) is invariant for any value of the effective 
coupling strength $P$ (since $a$ and $\k$ scales inversely). Also, the action corresponding 
to (\ref{eq:Ham}) is invariant and $\k$ flows to zero.

\section{Discussions}
We have shown that linear, higher spatial derivative theories 
give rise to a new kind of quantum phase transition. The change 
in the scaling of entanglement entropy occurs due to two effects:
\begin{enumerate}
\item {\it Kinematical:} Change in the density of states from increasing to 
constant. 
\item {\it Dynamical:} Excitation of larger $\ell$ modes compared to the 
canonical theory. 
\end{enumerate}
It is important to note that until now there is no fundamental understanding 
of the scaling of entanglement entropy with the density of states or the two-point 
correlation function. This work suggests a plausible link which will be explored 
in another work.

Recently, there has been interest in using entanglement as a probe to detect 
topological properties of many-body quantum states in two dimensional systems, 
like fractional quantum Hall effect \cite{2006-Kiteav.Preskill-PRL,2008-Li.Haldane-PRL}. 
Recently, it is found that Graphene shows an unconventional sequence of 
fractional quantum Hall states \cite{2012-Feldman.etal-Science}. Low-energy 
electronic states in Graphene are described by relativistic equation aka satisfying 
linear dispersion relation \cite{2009-Neto.etal-RMP}. The higher-order derivative term 
($b_2 \neq 0, b_3 = 0$) corresponds to Fermi gas has a constant density of 
state. Our work predicts that the above 2-D model will have an quantum 
phase transition. Currently, we are investigating the entanglement entropy to 
relate to the topological order.

It is usually expected that the higher derivative terms tame the UV
divergences.  In the case of two coupled harmonic oscillators
\cite{1993-Srednicki-PRL}, it is known that the entropy diverges when
the interaction strength is large. The infinitesimal boundary tends to
increase the interaction strength leading to the divergence of the
entanglement entropy \cite{1994-Bekenstein-Arx}. As it is clear from
our results, the divergence in the entanglement entropy can not be
resolved by introducing higher derivative terms.

Our results have interesting implications to black-hole entropy:
Numerical results for semi-classical black-holes have shown that close
to $90\%$ of Hawking radiation is in s-wave \cite{1976-Page-PRD}.  Our
results show that this is the case for the linear dispersion relation
(or large black-hole limit). However large $\ell$ modes contribute,
when the non-linear dispersion relations are dominant. It is also
important to note that earlier analyzes of Trans-Planckian effects on
the Hawking radiation have been performed for $\ell = 0$
\cite{1995-Unruh-PRD,1996-Corley.Jacobson-PRD}.  In the light of our
analysis those calculations have to be revisited.

Due to Hawking radiation, the mass of the black-holes will
decrease. As the size of the black-hole decreases, the curvature of
the event-horizon increases and hence one need to include higher
curvature terms
\cite{1993-Wald-PRD,1993-Jacobson.Myers-PRL,2008-Sen-GRG,2009-Kothawala.Padmanabhan-PRD}. Our
analysis shows that as the size of the black-hole decreases, below a
critical radius, the scaling of entropy changes instantaneously from
area to inverse of area. It is interesting to look at the plausible
implications of our result for the final stages of a microscopic
black-hole.

Our results also seem to be related to causal dynamical triangulation
approach of Ambj\'orn et al \cite{2005-Ambjorn.etal-PRL}.  There it
has been shown that the continuum limit corresponds to the linear
dispersion relation at large scales which in the UV limit changes to
anisotropic scaling of space (with respect to time) corresponding to
the dispersion relation $F(k) \propto k^3$.  Recently, Horava
\cite{2009-Horava-PRL} showed that gravity with dynamical critical
exponent $m = 3$ is perturbatively renormalizable (see also,
\cite{2009-Visser-PRD}). Our analysis suggests that causal dynamical
triangulation and Horava-Lifshitz model should also see a phase transition. 
These are under investigation.

\section*{Acknowledgments}
The authors wish to thank
J.~K.~Bhattacharjee, S.~Braunstein, S.~Das, D.~Jaiswal-Nagar, P.~Majumdar, J.~Mitra, 
R.~Narayanan, V.~Pai, J.~Samuel, D.~Sen, K.~Sengupta, S.~ Sinha, R.~Sorkin, 
A.~Taraphder and R.~Tibrewala for discussions and comments.  A special thanks to Subodh Shenoy for
demystifying some puzzling issues during the course of this work.  The
work is supported by the DST, Government of India through Ramanujan
fellowship and Max Planck-India Partner Group on Gravity and
Cosmology.

\begin{center}
\section*{Appendix}
\end{center}

\appendix

\section{Entanglement entropy from density matrix}

The Hamiltonian of a scalar field with modified dispersion relation,
Eq. (\ref{eq:Ham}), in $(3+1)$-dimensional flat space, is given
by:
\ba
H &=& \f{1}{2} \int d^3x \le[ \p^2(x) + |\vec\nabla\varphi(\vec x) |^2   + \f{\epsilon}{\kappa^2}(\vec\nabla^2\varphi(\vec x) )^2 + \f{\tau}{\kappa^4}|\vec\nabla^3\varphi(\vec x) |^2 \ri]~ . \la{Ham}
\ea
where $\kappa$ is the inverse length scale in the theory, $\epsilon =
0, \pm 1$ and $\tau = 0, 1$.

Decomposing the field and its conjugate momentum in partial waves
\ba
\varphi(\vec r)  = \sum_{lm} \f{\varphi_{lm}(r)}{r}~Y_{lm} (\theta,\phi)~,~ 
~\pi(\vec r)  = \sum_{lm} \f{\pi_{lm}(r)}{r}~Y_{lm} (\theta,\phi) \nn
\ea
yields:
%
\small{
\ba
H &=& \sum_{lm} H_{lm} \nn\\
&=&\sum_{lm} \f{1}{2} \int_0^\infty dr
\le[\p_{lm}^2(r) + r^2 \left( \f{\pa}{\pa r} \le( \f{\varphi_{lm} (r)}{r}\ri) \ri)^2 + \f{l(l+1)}{r^2} \varphi_{lm}^2(r) + \f{\epsilon}{\kappa^2} \le\{\f{d^2\varphi}{dr^2} - \f{l(l+1)}{r^2} \varphi_{lm}(r) \ri\}^2 \ri. \nn \\
&& \le.+ \f{\tau}{\kappa^4} \le\{ \le(\f{d^3\varphi}{dr^3} - \f{1}{r}\f{d^2\varphi}{dr^2} - \f{l(l+1)}{r^2}\f{d\varphi}{dr} + 3l(l+1)\f{\varphi}{r^3} \ri)^2 + \f{l(l+1)}{r^2} \le(\f{d^2\varphi}{dr^2}  - l(l+1) \f{\varphi}{r^2}\ri)^2\ri\}\right] 
\la{ham2}
\ea
}

The discretizing scheme used in this
work is different compared to that used by
Srednicki~\cite{1993-Srednicki-PRL}. Srednicki's mid-point discretization is not
suited in the presence of higher-derivative terms. In this work, we
have used central difference scheme, i.e., 
\ba f'(x) &=& \f{f(x+1) - f(x-1)}{2\triangle x}, \nn\\ 
   f''(x) &=& \f{f(x+1) - 2f(x) + f(x-1)}{(\triangle x)^2}, \nn\\ 
   f'''(x) &=& \f{f(x+2) - 2f(x+1) + 2f(x-1) - f(x-2)}{2(\triangle x)^3},
\ea 
which is 2nd order accurate.

Eq. (\ref{ham2}) can be written as the Hamiltonian of a set of $N$
coupled harmonic oscillators 
\be 
H = \sum_{j} H_{j} = \frac{1}{2a}
\sum_{i,j}^{N} \delta_{ij} \pi_{j}^2 + \vph_{j} \, K_{ij} \, \vph_i
\ee
where the off-diagonal elements of the matrix $K_{ij}$ represent the
interactions:
%
\ba
K_{ij} &=& \le[\f{1}{i^2} + \f{\epsilon}{a^2\kappa^2} + \f{\tau}{a^4\kappa^4} \le\{F_3^2(2) + \f{l(l+1)}{4} + \f{1}{4}\ri\}\ri]\delta_{i1}\delta_{j1} 
 - \f{\tau}{4a^4\kappa^4} \le[ \delta_{i2}\delta_{j2} + \delta_{i(N-1)}\delta_{j(N-1)} \ri] \nn \\ &&
+ \le[\f{(N-1)^2}{4N^2} + \f{\epsilon}{a^2\kappa^2} + \f{\tau}{a^4\kappa^4} \le\{F_1^2(N-1) + \f{l(l+1)}{(N-1)^2} + \f{1}{4} \ri\} \ri]  \delta_{iN}\delta_{jN} \nn \\ &&
+ \le[\f{(i^2+1)}{2i^2} + \f{2\epsilon}{a^2\kappa^2} + \f{\tau}{a^4\kappa^4} \le\{\f{1}{2} + F_2^2(i-1) + F_4^2(i+1) + \f{2(i^2+1)}{(i^2-1)^2}l(l+1) \ri\} \ri]\delta_{ij(i\neq 1,N)} \nn \\&& 
+ \le[ \f{l(l+1)}{i^2} + \f{\epsilon}{a^2\kappa^2} \le\{2 + \f{l(l+1)}{i^2} \ri\}^2 + \f{\tau}{a^4\kappa^4} \le\{F_2^2(i) + \f{l(l+1)}{i^2}F_4^2(i) \ri\} \ri] \delta_{ij} \nn\\&&
+ \le[- \f{\epsilon}{a^2\kappa^2} \le\{4 + l(l+1) \le(\f{1}{j^2} + \f{1}{(j+1)^2} \ri) \ri\}  + \f{\tau}{a^4\kappa^4} \le\{\f{F_1(j-1)}{2}(1-\delta_{j1}) + F_1(j)F_2(j) + \ri. \ri. \nn \\ && \le. \le. ~~~~~~~~~~~~  \f{l(l+1)}{j^2} F_4(j) + F_2(j+1)F_3(j+1) + \f{l(l+1)}{(j+1)^2} F_4(j+1) - \f{F_3(j+2)}{2}(1-\delta_{iN}) \ri\}\ri] \delta_{i,j+1} \nn \\&&
+ \le[- \f{\epsilon}{a^2\kappa^2} \le\{4 + l(l+1) \le(\f{1}{i^2} + \f{1}{(i+1)^2} \ri) \ri\}  + \f{\tau}{a^4\kappa^4} \le\{\f{F_1(i-1)}{2}(1-\delta_{i1}) + F_1(i)F_2(i) + \ri. \ri. \nn \\ && \le. \le. ~~~~~~~~~~~~ \f{l(l+1)}{i^2} F_4(i) + F_2(i+1)F_3(i+1) + \f{l(l+1)}{(i+1)^2} F_4(i+1) - \f{F_3(i+2)}{2} (1-\delta_{jN})\ri\} \ri] \delta_{i,j-1} \nn \\&&
-\le[\f{(j+1)^2}{4j(j+2)} - \f{\epsilon}{a^2\kappa^2} - \f{\tau}{a^4\kappa^4} \le\{F_1(j+1)F_3(j+1) + \f{l(l+1)}{(j+1)^2} + \f{F_2(j) - F_2(j+2)}{2} \ri\} \ri]\delta_{i,j+2} \nn \\ &&
- \le[\f{(i+1)^2}{4i(i+2)} - \f{\epsilon}{a^2\kappa^2} - \f{\tau}{a^4\kappa^4} \le\{F_1(i+1)F_3(i+1) + \f{l(l+1)}{(i+1)^2} + \f{F_2(i) - F_2(i+2)}{2} \ri\} \ri]\delta_{i,j-2} \nn \\ &&
+ \f{\tau}{a^4\kappa^4} \le[ \le\{\f{F_3(j+1) - F_1(j+2)}{2} \ri\} \delta_{i,j+3} + \le\{\f{F_3(i+1) - F_1(i+2)}{2} \ri\} \delta_{i,j-3} - \f{1}{4} (\delta_{i,j+4} + \delta_{i,j-4})\ri] 
 \la{kij}
\ea
where
\ba
F_1(i) &=& - 1 - \f{1}{i} - \f{l(l+1)}{2i^2}, F_2(i) = \f{2}{i} + \f{3l(l+1)}{i^3}, ~~~~\nn \\ 
 F_3(i) &=& 1 - \f{1}{i} + \f{l(l+1)}{2i^2}, F_4(i) = -2 - \f{l(l+1)}{i^2}.
\ea

Note that $K_{ij}$ contains nearest-neighbor (nn), next to nn (nnn), next to nnn (nnnn) and next to nnnn (nnnnn) interaction terms due to the presence of higher derivative terms in (\ref{ham2}). 
Schematically, 
{\small
\ba
K_{ij}  = \le( \begin{array}{lllllllll} 
{\times} & {\times} & {\times}& {\times}& {\times} & {} & {} & {} & {} \\
{\times} & {\times} & {\times} & {\times}& {\times} & {\times} & {} & {} & {}\\
{\times}  & {\times} & {\times} & {\times} & {\times} & {\times} & {\times} & {} & {}\\
{\times} & {\times}  & {\times} & {\times} & {\times} & {\times} & {\times}& {\times} & {} \\
{\times} & {\times} & {\times}  & {\times} & {\times} & {\times} & {\times} & {\times} & {\times} \\
{} & {\times} & \times{} & {\times}  & {\times} & {\times} & {\times} & {\times} & {\times} \\
{} & {} & {\times} & {\times} & {\times}  & {\times} & {\times}  & {\times} & \times{}\\
{} & {} & {} & {\times} & {\times}  & {\times} & {\times}  & {\times} & {\times}\\
{} & {} & {} & {} & {\times}  & {\times} & {\times}  & {\times} & {\times}
\la{mat1} \\ 
\end{array} \ri) 
\ea
}
\label{disc-SchHam}
A brief description of how to calculate entropy from the above
Hamiltonian is the following.  The density matrix, tracing over the
first $n$ of $N$ oscillators ($r \equiv r_{n+1},\dots, r_N$), is given
by:
%
\be
\rho_{\rm reduced} = 
\int \prod_{i=1}^n~dr_i~\varphi (r_1,\dots,r_n;r_{n+1},\dots,r_{N})~ 
\varphi^\star (r_1,\dots,r_n;r'_{n+1},\dots,r_{N}') 
\la{den2} 
\ee
%
where $r$ and $r'$ represent radial distances of the points, outside the horizon, from the center.
The ground state is
\be
\phi(r_1,\dots,r_N) = \prod_{i=1}^N N_i 
\exp( -\f{1}{2} k_{Di}^{\frac{1}{2}} {\underbar r}_i^2), ~~~~~
\la{gs2}
\ee
(where $\underbar r = Ur~,~UKU^T= K_D~$ a diagonal matrix) the
corresponding density matrix (\ref{den2}) can be evaluated exactly
(the superscript $(0)$ signifies GS):
%
\be
\r_{\rm reduced}^{(0)}  \sim \exp\le[ -(r^T \gamma r + r'^T\gamma r')/2 + r^T\beta r \ri] 
\la{gsden}
\ee
where:
\be 
\O \sim K^{1/2} = \le( \begin{array}{ll} 
A& B  \\ 
B^T & C \\
\end{array} \ri)~,~\beta=\f{1}{2}B^TA^{-1}B~,~\gamma=C-\beta~.
\ee
Note that $B$ and $\beta$ are non-zero if and only if there are
interactions. The Gaussian nature of the above density matrix lends
itself to a series of diagonalisations
[$V\c V^T =$ diag, ${\bar\b} \equiv \c_D^{-\f{1}{2}} V \b V^T \c_D^{-\f{1}{2}}$, 
$W {\bar \b} W^T = $ diag, $v_i \in v \equiv W^T (V \c V^T)^{\f{1}{2}} VT$], 
such that it reduces to a product of $(N-n)$, $2$-oscillator density
matrices, in each of which one oscillator is traced over \cite{1993-Srednicki-PRL}:
\be
\rho^{(0)}_{_{\rm reduced}}  \sim
\prod_{i=1}^{N-n} 
\exp\le[-\f{v_i^2+v_i'^2}{2} + {\bar \beta}_i v_i v_i' \ri] ~.
\ee
The corresponding entropy is given by:
\be
S = \sum_{i=1}^{N-n} 
\le( - \ln[1-\xi_i] - \f{\xi_i}{1-\xi_i}\ln\xi_i  \ri) ~
\le[
\xi_i = \f{{\bar \b}_i}{1+ \sqrt{1-{\bar \beta}_i^2}} \ri]~.
\ee
Thus, for the full Hamiltonian $H=\sum_{lm}H_{lm}$, the entropy is:
\be
S = \sum_{l=0}^{\infty} (2 l+ 1) S_l 
\ee
where the degeneracy factor $(2l+1)$ follows from spherical symmetry
of the Hamiltonian. In practice, we will replace the upper bound of
the sum in the above to a large value $l_{max}$. For the interaction
matrix without any correction i.e. with $\epsilon = \tau = 0$ in
(\ref{kij}), the above entropy, computed numerically, turned out to be
\cite{1986-Bombelli.etal-PRD,1993-Srednicki-PRL}:
\be
S = 0.3 (n+1/2)^2 \equiv  0.3 \le(\f{R}{a}\ri)^2~.
\ee

\section{Asymptotic analysis: Convergence of entropic contribution}

For $l>>j$, the Hamiltonian can be written as
\be
H \sim \f{1}{2a} \sum_{i} \pi_i^2 + \f{l^2}{i^2} \left(1 + \epsilon P ~\f{l^2}{i^2} + \tau P^2 ~\f{l^4}{i^4}\right)\phi_i^2 + \mbox{interaction/perturbative terms...}\label{eq:Hamil2}
\ee

In the case of $\e=1$ and $\t=0$ in the Hamiltonian (\ref{eq:Hamil2}),
we get $\xi(P\rightarrow 0) = \f{n^4}{64l^4}$ (where, $P =
\f{1}{a^2\kappa^2}$), where as the corresponding result of Srednicki
was $\f{n^4}{16l^4}$ \cite{1993-Srednicki-PRL} and $\xi(P\rightarrow 1) =
\f{n^4}{l^4}$.

In the case of $\e=0$ and $\t=1$, we get $\xi(P\rightarrow 0) =
\f{n^4}{l^4}$ and $\xi(P\rightarrow 1) = \f{17n^4}{16l^4} +
\f{n^6}{l^6} + \f{n^{12}}{l^{12}}$. These values confirm that the
entropy converges for large $l$ and justifies the use of an upper
cutoff $l_{max}$ for numerical estimation.

\section{Calculation of correlation function}

The two point correlation function of the fields also provide
information about structure of the scalar field from the linear to
non-linear dispersion relations.
The (equal time) two-point correlation function or the Wightman
function is given by
\ba
G^+ (x^{\mu}, y^{\mu}) &=& \int \f{d^3k}{(2\pi)^3} \f{\exp[i\vec k.(\vec x - \vec y)]}{2\omega}  \nn \\
&=& \f{1}{4\pi^2r} \int_0^\infty \f{k}{\omega}~ \sin(kr)~ dk  %
 \label{eq:wightman1}
\ea
where $r = |{\vec x}-{\vec y}|$. Eq. (\ref{eq:wightman1}) can be rewritten for a general dispersion relation $\omega = k^m$ and with substitution $kr=\alpha$ as
\be
G^+ (x^{\mu}, y^{\mu}) = \f{r^{m-3}}{4\pi^2} \int_0^\infty \f{\sin \alpha}{\alpha^{m-1}}~ d\alpha  %
 \label{eq:wightman2}
\ee

Note that for linear ($m=1$) and quadratic ($m=2$) dispersion models correlation {\it decays} with increasing distance which explains the area-law behavior of the entropy. Interestingly when the {\it third} order correction ($m=3$) is dominant, the correlation function essentially becomes {\it scale invariant}.

\section{Entropy per partial wave for different dispersion models}

\begin{figure}[!htb]
\begin{minipage}{18pc}
\includegraphics[width=18pc]{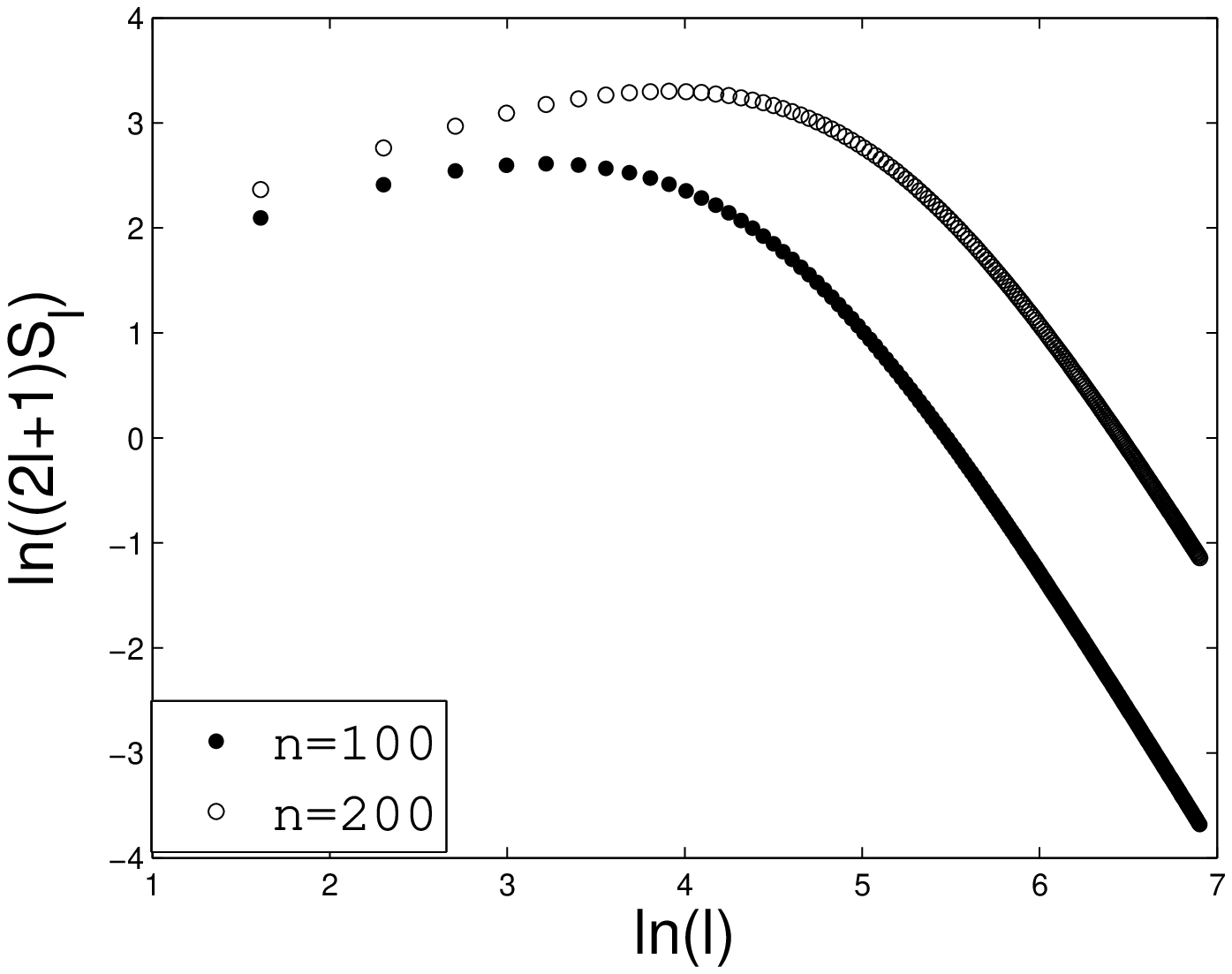}
\caption{\label{fig:1storder}Entropy distribution per partial wave $(2 \ell + 1) S_{\ell}$ and $\ell$, in linear dispersive system ($m=1$), for $n = 200$ and $n=100$.}
\end{minipage}\hspace{1pc}%
\begin{minipage}{18pc}
\includegraphics[width=18pc]{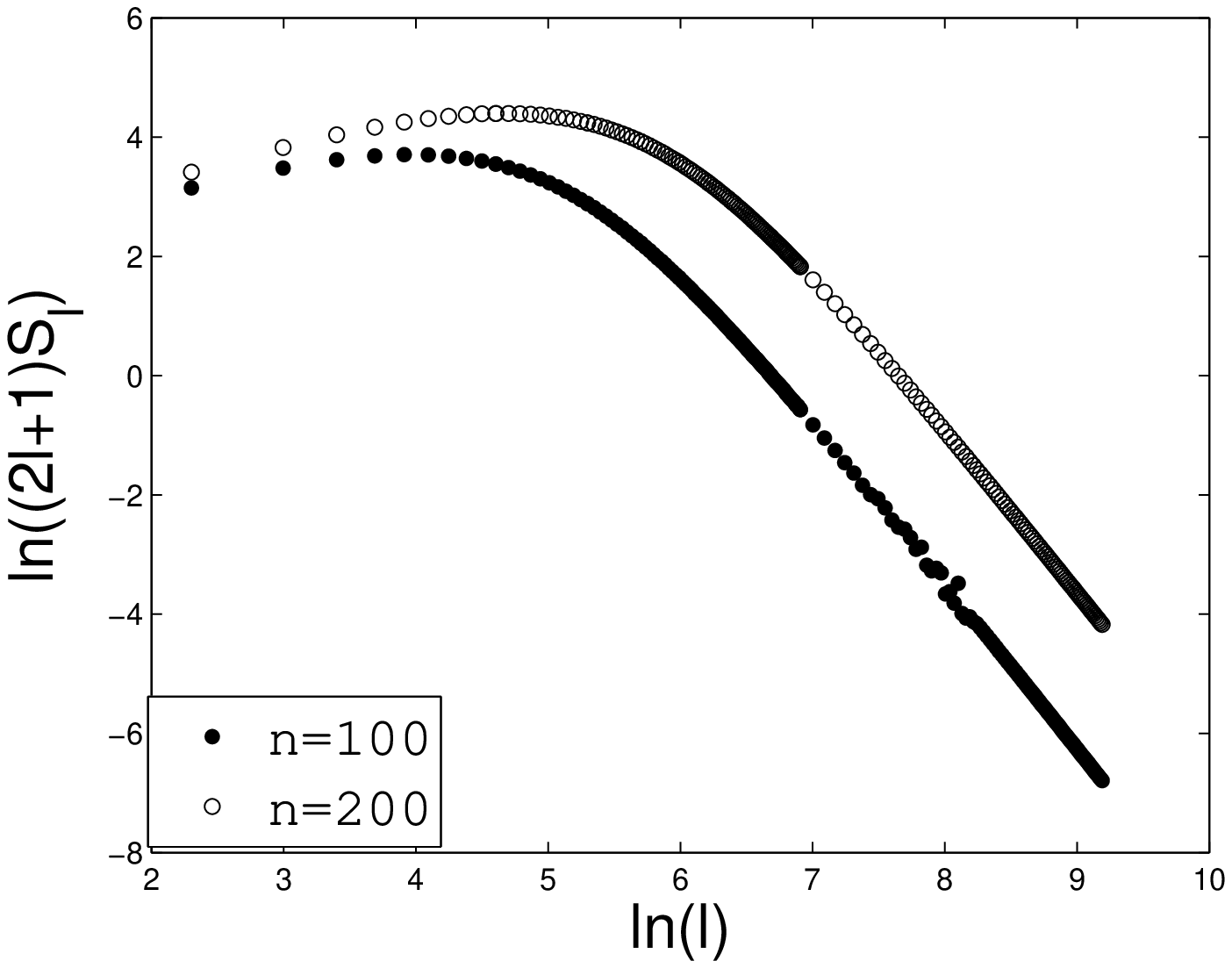} 
\caption{\label{fig:2ndorder-App}Entropy distribution per partial wave $(2 \ell + 1) S_{\ell}$ and $\ell$, in quadratic dispersive system ($m=2$), for $n = 200$ and $n=100$.}
\end{minipage}\hspace{1pc}
\label{fig:1st&2ndorder}
\end{figure}

\begin{figure}[!htb]
\includegraphics[width=40pc]{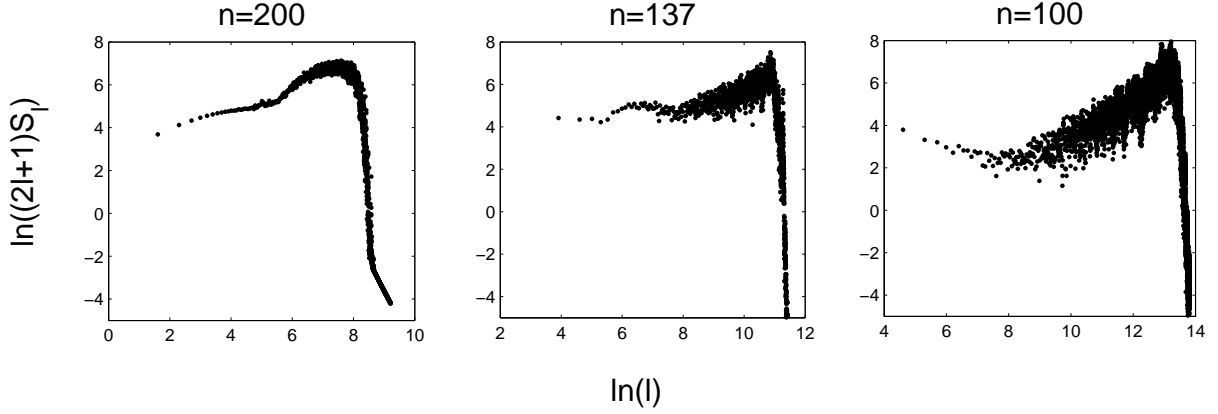}
\caption{Entropy distribution per partial wave $(2\ell + 1) S_{\ell}$ and $\ell$, in cubic dispersive system ($m=3$), for $n = 200$, $n=137$ and $n=100$.}
\label{fig:3rdorder-App}
\end{figure}
Fig. \ref{fig:1storder} and Fig. \ref{fig:2ndorder-App} shows that in linear and quadratic dispersive theories, with decreasing $n$, entropy per partial wave decreases. On the other hand for cubic dispersion theory exactly opposite behavior is found (see Fig. \ref{fig:3rdorder-App}).

\section{Particle in a box model}

The Schr\"odinger equation corresponding to a general non-linear
dispersion model is:
\be 
\f{d^{2m}\psi}{dx^{2m}} + {\cal E} \psi = 0 \qquad \quad {\cal E} = E \kappa_{{}_P}^{2(m-1)} . \label{eq:Shrodinger}
\ee 
The most general solution is given by
\be
\psi(x) = \sum_{j = 1}^{2m} C_j\exp[(-1)^{\f{2j-1}{2m}} {\cal E}^{\f{1}{2m}} x] \, ,
\ee
where $C_j$'s are constants to be determined by the boundary
conditions.  In particular, for $m = 1$, $2$ and $3$, the general
solution is given by:
\ba
\psi_{m=1} &=& c_1 \cos({\cal E}^{\f{1}{2}} x) + c_2 \sin ({\cal E}^{\f{1}{2}} x) \\
\psi_{m=2} &=& \exp\left(\f{{\cal E}^{\f{1}{4}} x}{\sqrt{2}}\right) \left[c_1\cos\left(\f{{\cal E}^{\f{1}{4}} x}{\sqrt{2}}\right) + c_2 \sin \left(\f{{\cal E}^{\f{1}{4}} x}{\sqrt{2}}\right) \right] \nn \\
&&+ \exp\left(-\f{{\cal E}^{\f{1}{4}} x}{\sqrt{2}}\right) \left[c_3\cos\left(\f{{\cal E}^{\f{1}{4}} x}{\sqrt{2}}\right) + c_4 \sin \left(\f{{\cal E}^{\f{1}{4}} x}{\sqrt{2}}\right) \right]\\
\psi_{m=3} &=& \exp\left(\f{{\sqrt{3}\cal E}^{\f{1}{6}} x}{2}\right) \left[c_1\cos\left(\f{{\cal E}^{\f{1}{6}} x}{2}\right) + c_2 \sin \left(\f{{\cal E}^{\f{1}{6}} x}{2}\right) \right] \nn \\
&&+ \exp\left(-\f{\sqrt{3}{\cal E}^{\f{1}{6}} x}{2}\right) \left[c_3\cos\left(\f{{\cal E}^{\f{1}{6}} x}{2}\right) + c_4 \sin \left(\f{{\cal E}^{\f{1}{6}} x}{2}\right) \right]\nn\\
&& + c_5 \cos\left({\cal E}^{\f{1}{6}} x\right) + c_6 \sin \le({\cal E}^{\f{1}{6}} x\ri)
\ea
For a particle in a box model, only exponentially growing and decaying
solutions exist for even $m$, while for odd $m$ one has stationary
solutions [$\sim \sin ({\cal E}^{\f{1}{2m}} x)$].  With the
appropriate boundary conditions at $x = 0$ and $L$, one finds the
energy eigenvalues are $E_{\nu}$:
{\small
\be
E_\nu \sim \left(\f{\nu \pi}{L} \kappa^{\f{1-m}{m}} \right)^{2m} \qquad \nu = 1,2,.. \label{eq:eigen}
\ee
}
which leads to:
{\small
\be
\frac{\mbox{\small Ground state energy eigenvalue for $m=1$}}{\mbox{\small Ground state energy eigenvalue for $m>1$}}
= \left[\frac{L \kappa}{\pi}\right]^{2 (m - 1)} \label{eq:eigen-ratio}
\ee
}

\begin{figure}[!htb]
\begin{center}
\includegraphics[scale=0.30]{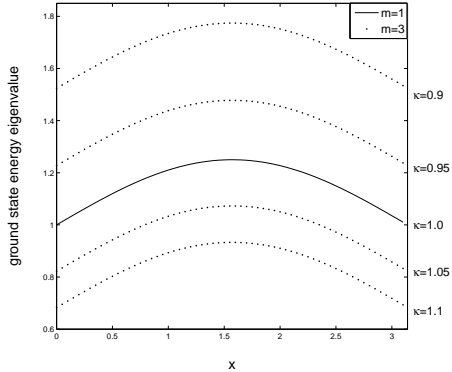}
\caption{Energy eigenvalues for $m = 1$ and $3$, $L = \pi$ and varying $\kappa$.}
\label{fig:particleinabox}
\end{center}
\end{figure}
For $m > 1$, with decreasing $\kappa$, the ground state energy
eigenvalue of the system increases.  Fig. (\ref{fig:particleinabox})
shows that at $\kappa=\pi/L$, it crosses the ground state energy
eigenvalue for the system satisfying linear dispersion relation
($m=1$) and becomes more than that for $\kappa<\pi/L$.

Eq. (\ref{eq:eigen-ratio}) implies that, for $\kappa <\pi/L$, the
ground state energy eigenvalue in the non-linear dispersion theory is
higher compared to that of linear dispersion theory.  As in the field
theory model, where there is a cross-over from the linear to
non-linear regime, with increasing $P$ (or decreasing $\kappa$), for
$P>P_c$, the system needs to readjust in such a way that the ground
state energy of the system increases. In other words, the cross-over
of the dispersion relation catalyzes \textit{larger population} of
higher energy quantum states compared to low-energy states.

\section{Transition in the thermodynamic limit}

\begin{figure}[!htb]
\begin{center}
\includegraphics[scale=0.6]{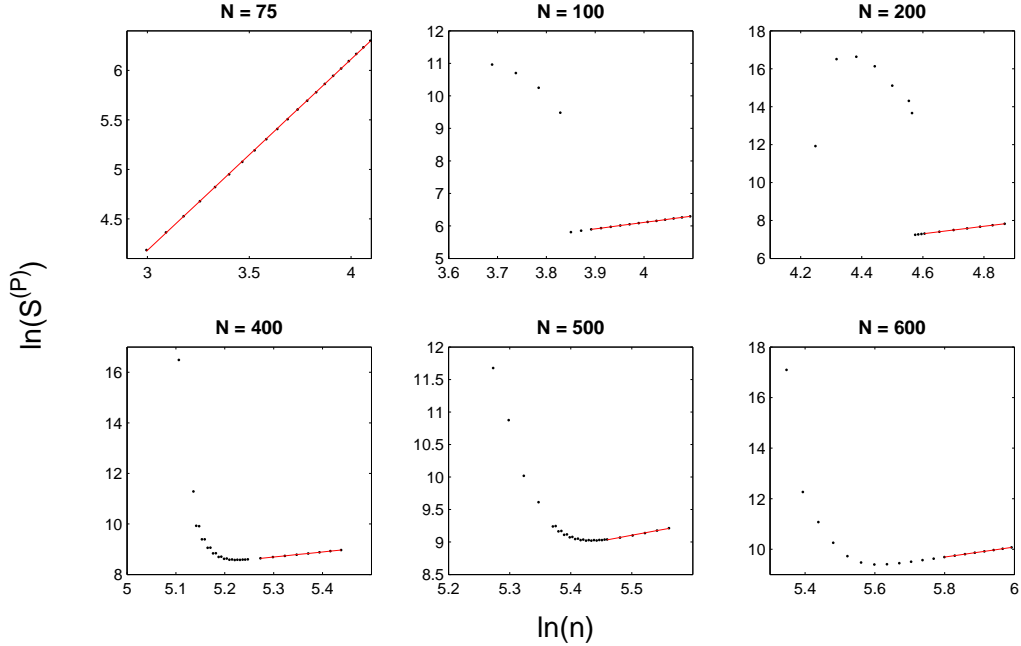}
\caption{Entropy profile for $P=10^{-5}$ with different system size $N$.}
\label{fig:q5_varyingN}
\end{center}
\end{figure}

Fig. \ref{fig:q5_varyingN} shows that, for fixed $P$ with increasing
$N$ the transition happens approximately at the same $n_c/N$ though it looses
it's sharpness. The inverse scaling does not appear for $N=75$ implying 
that $P_c$ decreases with increasing $N$.

\vskip 0.16truein


\begin{thebibliography}{22}
\expandafter\ifx\csname natexlab\endcsname\relax\def\natexlab#1{#1}\fi
\expandafter\ifx\csname bibnamefont\endcsname\relax
  \def\bibnamefont#1{#1}\fi
\expandafter\ifx\csname bibfnamefont\endcsname\relax
  \def\bibfnamefont#1{#1}\fi
\expandafter\ifx\csname citenamefont\endcsname\relax
  \def\citenamefont#1{#1}\fi
\expandafter\ifx\csname url\endcsname\relax
  \def\url#1{\texttt{#1}}\fi
\expandafter\ifx\csname urlprefix\endcsname\relax\def\urlprefix{URL }\fi
\providecommand{\bibinfo}[2]{#2}
\providecommand{\eprint}[2][]{\url{#2}}

\bibitem[{\citenamefont{{Horodecki} et~al.}(2009)\citenamefont{{Horodecki},
  {Horodecki}, {Horodecki}, and {Horodecki}}}]{2009-Horodecki.etal-RMP}
\bibinfo{author}{\bibfnamefont{R.}~\bibnamefont{{Horodecki}}},
  \bibinfo{author}{\bibfnamefont{P.}~\bibnamefont{{Horodecki}}},
  \bibinfo{author}{\bibfnamefont{M.}~\bibnamefont{{Horodecki}}},
  \bibnamefont{and}
  \bibinfo{author}{\bibfnamefont{K.}~\bibnamefont{{Horodecki}}},
  \bibinfo{journal}{Rev. Mod. Phys.} \textbf{\bibinfo{volume}{81}},
  \bibinfo{pages}{865} (\bibinfo{year}{2009}).

\bibitem[{\citenamefont{Eisert et~al.}(2010)\citenamefont{Eisert, Cramer, and
  Plenio}}]{2010-Eisert.etal-RMP}
\bibinfo{author}{\bibfnamefont{J.}~\bibnamefont{Eisert}},
  \bibinfo{author}{\bibfnamefont{M.}~\bibnamefont{Cramer}}, \bibnamefont{and}
  \bibinfo{author}{\bibfnamefont{M.~B.} \bibnamefont{Plenio}},
  \bibinfo{journal}{Rev. Mod. Phys.} \textbf{\bibinfo{volume}{82}},
  \bibinfo{pages}{277} (\bibinfo{year}{2010}), \eprint{0808.3773}.

\bibitem[{\citenamefont{Bombelli et~al.}(1986)\citenamefont{Bombelli, Koul,
  Lee, and Sorkin}}]{1986-Bombelli.etal-PRD}
\bibinfo{author}{\bibfnamefont{L.}~\bibnamefont{Bombelli}},
  \bibinfo{author}{\bibfnamefont{R.~K.} \bibnamefont{Koul}},
  \bibinfo{author}{\bibfnamefont{J.-H.} \bibnamefont{Lee}}, \bibnamefont{and}
  \bibinfo{author}{\bibfnamefont{R.~D.} \bibnamefont{Sorkin}},
  \bibinfo{journal}{Phys. Rev.} \textbf{\bibinfo{volume}{D34}},
  \bibinfo{pages}{373} (\bibinfo{year}{1986}).

\bibitem[{\citenamefont{Srednicki}(1993)}]{1993-Srednicki-PRL}
\bibinfo{author}{\bibfnamefont{M.}~\bibnamefont{Srednicki}},
  \bibinfo{journal}{Phys. Rev. Lett.} \textbf{\bibinfo{volume}{71}},
  \bibinfo{pages}{666} (\bibinfo{year}{1993}).

\bibitem[{\citenamefont{Das et~al.}(2008)\citenamefont{Das, Shankaranarayanan,
  and Sur}}]{2008-Das.etal-PRD}
\bibinfo{author}{\bibfnamefont{S.}~\bibnamefont{Das}},
  \bibinfo{author}{\bibfnamefont{S.}~\bibnamefont{Shankaranarayanan}},
  \bibnamefont{and} \bibinfo{author}{\bibfnamefont{S.}~\bibnamefont{Sur}},
  \bibinfo{journal}{Phys. Rev.} \textbf{\bibinfo{volume}{D77}},
  \bibinfo{pages}{064013} (\bibinfo{year}{2008}), \eprint{0705.2070}.

\bibitem[{\citenamefont{{Sondhi} et~al.}(1997)\citenamefont{{Sondhi}, {Girvin},
  {Carini}, and {Shahar}}}]{1997-Sondhi.etal-RMP}
\bibinfo{author}{\bibfnamefont{S.~L.} \bibnamefont{{Sondhi}}},
  \bibinfo{author}{\bibfnamefont{S.~M.} \bibnamefont{{Girvin}}},
  \bibinfo{author}{\bibfnamefont{J.~P.} \bibnamefont{{Carini}}},
  \bibnamefont{and} \bibinfo{author}{\bibfnamefont{D.}~\bibnamefont{{Shahar}}},
  \bibinfo{journal}{Rev. Mod. Phys.} \textbf{\bibinfo{volume}{69}},
  \bibinfo{pages}{315} (\bibinfo{year}{1997}).

\bibitem[{\citenamefont{Sachdev}(2001)}]{2001-05-30-SachdevSubir-Quantumphaset%
ransitions}
\bibinfo{author}{\bibfnamefont{S.}~\bibnamefont{Sachdev}},
  \emph{\bibinfo{title}{Quantum phase transitions}}, 
  (\bibinfo{publisher}{Cambridge University Press}, \bibinfo{year}{2001}), ISBN
  \bibinfo{isbn}{9780521004541}.

 \bibitem[{\citenamefont{Carr}(2011)}]{2011-Carr-QPT}
\bibinfo{author}{\bibfnamefont{L.}~\bibnamefont{Carr}},
  \emph{\bibinfo{title}{Understanding quantum phase transitions}}, 
  (\bibinfo{publisher}{CRC Press}, \bibinfo{year}{2011}), ISBN
  \bibinfo{isbn}{9781439802519}. 
    
  
\bibitem[{\citenamefont{{Osterloh} et~al.}(2002)\citenamefont{{Osterloh},
  {Amico}, {Falci}, and {Fazio}}}]{2002-Osterloh.etal-Nature}
\bibinfo{author}{\bibfnamefont{A.}~\bibnamefont{{Osterloh}}},
  \bibinfo{author}{\bibfnamefont{L.}~\bibnamefont{{Amico}}},
  \bibinfo{author}{\bibfnamefont{G.}~\bibnamefont{{Falci}}}, \bibnamefont{and}
  \bibinfo{author}{\bibfnamefont{R.}~\bibnamefont{{Fazio}}},
  \bibinfo{journal}{Nature} \textbf{\bibinfo{volume}{416}},
  \bibinfo{pages}{608} (\bibinfo{year}{2002}), \eprint{arXiv:quant-ph/0202029}.

\bibitem[{\citenamefont{{Osborne} and
  {Nielsen}}(2002)}]{2002-Osborne.Nielsen-PRA}
\bibinfo{author}{\bibfnamefont{T.~J.} \bibnamefont{{Osborne}}}
  \bibnamefont{and} \bibinfo{author}{\bibfnamefont{M.~A.}
  \bibnamefont{{Nielsen}}}, \bibinfo{journal}{Phys. Rev. A}
  \textbf{\bibinfo{volume}{66}}, \bibinfo{eid}{032110} (\bibinfo{year}{2002}),
  \eprint{arXiv:quant-ph/0202162}.

\bibitem[{\citenamefont{{Wu} et~al.}(2004)\citenamefont{{Wu}, {Sarandy}, and
  {Lidar}}}]{2004-Wu.etal-PRL}
\bibinfo{author}{\bibfnamefont{L.-A.} \bibnamefont{{Wu}}},
  \bibinfo{author}{\bibfnamefont{M.~S.} \bibnamefont{{Sarandy}}},
  \bibnamefont{and} \bibinfo{author}{\bibfnamefont{D.~A.}
  \bibnamefont{{Lidar}}}, \bibinfo{journal}{Phys. Rev. Lett.}
  \textbf{\bibinfo{volume}{93}}, \bibinfo{eid}{250404} (\bibinfo{year}{2004}),
  \eprint{arXiv:quant-ph/0407056}.

\bibitem[{\citenamefont{{Rieper} et~al.}(2010)\citenamefont{{Rieper}, {Anders},
  and {Vedral}}}]{2010-Rieper.etal-NJP}
\bibinfo{author}{\bibfnamefont{E.}~\bibnamefont{{Rieper}}},
  \bibinfo{author}{\bibfnamefont{J.}~\bibnamefont{{Anders}}}, \bibnamefont{and}
  \bibinfo{author}{\bibfnamefont{V.}~\bibnamefont{{Vedral}}},
  \bibinfo{journal}{New J. Phys.} \textbf{\bibinfo{volume}{12}},
  \bibinfo{pages}{025017} (\bibinfo{year}{2010}), \eprint{0908.0636}.

\bibitem{1972-Anderson-Science}
P.~W.~Anderson, Science {\bf 177}, 393-396 (1972)

\bibitem[{\citenamefont{Toledano}(1987)}]{1987-Toledano-Landau}
\bibinfo{author}{\bibfnamefont{J.~C.}~\bibnamefont{Toledano},
and \bibfnamefont{P}.~\bibnamefont{Toledano}},
  \emph{\bibinfo{title}{The Landau Theory of Phase Transitions}}, 
  (\bibinfo{publisher}{World Scientific}, \bibinfo{year}{1987}).
  
\bibitem[{\citenamefont{{Hornreich} et~al.}(1975)\citenamefont{{Honreich},
  {Luban}, and {Shtrikman}}}]{1975-Honreich.etal-PRL}
\bibinfo{author}{\bibfnamefont{R.~M}~\bibnamefont{{Honreich}}},
  \bibinfo{author}{\bibfnamefont{M.}~\bibnamefont{{Luban}}}, and
  \bibinfo{author}{\bibfnamefont{S.} \bibnamefont{{Shritman}}},
  \bibinfo{journal}{Phys. Rev. Lett.} \textbf{\bibinfo{volume}{35}},
  \bibinfo{eid}{1678} (\bibinfo{year}{1975}).  

\bibitem[{\citenamefont{{Unruh}}(1995)}]{1995-Unruh-PRD}
\bibinfo{author}{\bibfnamefont{W.~G.} \bibnamefont{{Unruh}}},
  \bibinfo{journal}{Phys. Rev.} \textbf{\bibinfo{volume}{D51}},
  \bibinfo{pages}{2827} (\bibinfo{year}{1995}).

\bibitem[{\citenamefont{Corley and Jacobson}(1996)}]{1996-Corley.Jacobson-PRD}
\bibinfo{author}{\bibfnamefont{S.}~\bibnamefont{Corley}} \bibnamefont{and}
  \bibinfo{author}{\bibfnamefont{T.}~\bibnamefont{Jacobson}},
  \bibinfo{journal}{Phys. Rev.} \textbf{\bibinfo{volume}{D54}},
  \bibinfo{pages}{1568} (\bibinfo{year}{1996}).

\bibitem[{\citenamefont{Padmanabhan}(1999)}]{1999-Padmanabhan-PRD}
\bibinfo{author}{\bibfnamefont{T.}~\bibnamefont{Padmanabhan}},
  \bibinfo{journal}{Phys. Rev.} \textbf{\bibinfo{volume}{D59}},
  \bibinfo{pages}{124012} (\bibinfo{year}{1999}).

\bibitem[{\citenamefont{Visser}(2009)}]{2009-Visser-PRD}
\bibinfo{author}{\bibfnamefont{M.}~\bibnamefont{Visser}},
  \bibinfo{journal}{Phys. Rev.} \textbf{\bibinfo{volume}{D80}},
  \bibinfo{pages}{025011} (\bibinfo{year}{2009}), \eprint{0902.0590}.

\bibitem[{\citenamefont{{Lookman} et~al.}(2003)\citenamefont{{Lookman},
  {Shenoy}, {Rasmussen}, {Saxena}, and {Bishop}}}]{2003-Lookman.etal-PRB}
\bibinfo{author}{\bibfnamefont{T.}~\bibnamefont{{Lookman}}},
  \bibinfo{author}{\bibfnamefont{S.~R.} \bibnamefont{{Shenoy}}},
  \bibinfo{author}{\bibfnamefont{K.~O.} \bibnamefont{{Rasmussen}}},
  \bibinfo{author}{\bibfnamefont{A.}~\bibnamefont{{Saxena}}}, \bibnamefont{and}
  \bibinfo{author}{\bibfnamefont{A.~R.} \bibnamefont{{Bishop}}},
  \bibinfo{journal}{Phys. Rev. B.} \textbf{\bibinfo{volume}{67}},
  \bibinfo{eid}{024114} (\bibinfo{year}{2003}).
    
\bibitem{2006-Kiteav.Preskill-PRL}    
A.~Kitaev and J.~Preskill, Phys. Rev. Letts.
{\bf 96}, 110404 (2006).

\bibitem{2008-Li.Haldane-PRL}
H. Li and F. D. M. Haldane, Phys. Rev. Letts. {\bf 101}, 010504 (2008).

\bibitem{2012-Feldman.etal-Science}
B. Feldman, B. Krauss, J. Smet  and A. Yacoby, 
Science, {\bf 337}, 1196-1199 (2012).

\bibitem{2009-Neto.etal-RMP}
A. H. Castro Neto, F. Guinea, N. M. R. Peres, K. S. Novoselov, and A. K. Geim
Rev. Mod. Phys. {\bf 81}, 109 (2009)
 
   
\bibitem{1994-Bekenstein-Arx} 
 J.~D.~Bekenstein,
  gr-qc/9409015.

\bibitem{1976-Page-PRD} 
  D.~N.~Page,
  Phys.\ Rev.\ D {\bf 13}, 198 (1976).




\bibitem[{\citenamefont{Wald}(1993)}]{1993-Wald-PRD}
\bibinfo{author}{\bibfnamefont{R.}~\bibnamefont{Wald}}, \bibinfo{journal}{Phys.
  Rev.} \textbf{\bibinfo{volume}{D48}}, \bibinfo{pages}{R3427}
  (\bibinfo{year}{1993}), \eprint{gr-qc/9307038}.

\bibitem[{\citenamefont{Jacobson and Myers}(1993)}]{1993-Jacobson.Myers-PRL}
\bibinfo{author}{\bibfnamefont{T.}~\bibnamefont{Jacobson}} \bibnamefont{and}
  \bibinfo{author}{\bibfnamefont{R.~C.} \bibnamefont{Myers}},
  \bibinfo{journal}{Phys. Rev. Lett.} \textbf{\bibinfo{volume}{70}},
  \bibinfo{pages}{3684} (\bibinfo{year}{1993}), \eprint{hep-th/9305016}.

\bibitem[{\citenamefont{Sen}(2008)}]{2008-Sen-GRG}
\bibinfo{author}{\bibfnamefont{A.}~\bibnamefont{Sen}}, \bibinfo{journal}{Gen.
  Rel. Grav.} \textbf{\bibinfo{volume}{40}}, \bibinfo{pages}{2249}
  (\bibinfo{year}{2008}), \eprint{0708.1270}.

\bibitem[{\citenamefont{Kothawala and
  Padmanabhan}(2009)}]{2009-Kothawala.Padmanabhan-PRD}
\bibinfo{author}{\bibfnamefont{D.}~\bibnamefont{Kothawala}} \bibnamefont{and}
  \bibinfo{author}{\bibfnamefont{T.}~\bibnamefont{Padmanabhan}},
  \bibinfo{journal}{Phys. Rev.} \textbf{\bibinfo{volume}{D79}},
  \bibinfo{pages}{104020} (\bibinfo{year}{2009}), \eprint{0904.0215}.


\bibitem[{\citenamefont{Ambjorn} et~al.} (2005)
         \citenamefont{{Jurkiewicz}, and {Loll}}]{2005-Ambjorn.etal-PRL}
\bibinfo{author}{\bibfnamefont{J.}~\bibnamefont{Ambjorn}},
\bibinfo{author}{\bibfnamefont{J.}~\bibnamefont{Jurkiewicz}},
\bibinfo{author}{\bibfnamefont{R.}~\bibnamefont{Loll}},
  \bibinfo{journal}{Phys. Rev. Lett.} \textbf{\bibinfo{volume}{95}},
  \bibinfo{pages}{171301} (\bibinfo{year}{2005}).

\bibitem[{\citenamefont{Horava}(2009)}]{2009-Horava-PRL}
\bibinfo{author}{\bibfnamefont{P.}~\bibnamefont{Horava}},
  \bibinfo{journal}{Phys. Rev. Lett.} \textbf{\bibinfo{volume}{180}},
  \bibinfo{pages}{161301} (\bibinfo{year}{2009}).

\end{thebibliography}
\end{document}